\documentclass[aps,preprint,nofootinbib]{revtex4}%
\usepackage{amsfonts}
\usepackage{amsmath}
\usepackage{amssymb}
\usepackage{graphicx}%
\providecommand{\U}[1]{\protect\rule{.1in}{.1in}}

\begin{document}
\preprint{ }
\title[Short title for running header]{An Analysis of the First Order Form of Gauge Theories}
\author{N. Kiriushcheva}
\email{nkiriush@uwo.ca}
\affiliation{The Department of Applied Mathematics, The University of Western Ontario,
London, Ontario, N6A 5B7, Canada}
\author{S. V. Kuzmin}
\email{skuzmin@uwo.ca}
\affiliation{The Department of Applied Mathematics, The University of Western Ontario,
London, Ontario, N6A 5B7, Canada}
\author{D. G. C. McKeon}
\email{dgmckeo2@uwo.ca}
\affiliation{The Department of Applied Mathematics, The University of Western Ontario,
London, Ontario, N6A 5B7, Canada}
\affiliation{The Department of Mathematics and Compuer Science, Algoma University, Sault
Ste. Matie, Ontario, P6A 2G4, Canada}

\begin{abstract}
The first order form of a Maxwell theory and $U(1)$ gauge theory in which a
gauge invariant mass term appears is analyzed using the Dirac procedure. The
form of the gauge transformation which leaves the action invariant is derived
from the constraints present. A non-Abelian generalization is similarly
analyzed. This first order three dimensional massive gauge theory is rewritten
in terms of two interacting vector fields. The constraint structure when using
light-cone coordinates is considered. The relationship between first and
second order forms of the two-dimensional Einstein-Hilbert action is explored
where a Lagrange multiplier is used to ensure their equivalence.

\end{abstract}
\volumeyear{year}
\volumenumber{number}
\issuenumber{number}
\eid{identifier}
\date{\today}
\received{}

\maketitle

\section{Introduction}

In gauge theories, superfluous degrees of freedom are retained in the action
to ensure that invariances present in the theory are manifest. Both the
Yang-Mills (YM) and Einstein-Hilbert (EH) gauge actions can be written in both
first and second order forms. Most calculations have been performed using the
second order form. This form has the advantage of employing fewer fields than
the first order formalism. However, in the first order formalism, the vertices
are considerably simpler than in the second order formalism. For YM theory,
this means that the complicated three-point and four-point vertices for the
vector field are replaced by a relatively simple vertex involving two vector
fields and one field strength [26]. In the case of the EH action, a
complicated non-polynomial iteration for the fluctuations about a background
metric is replaced by a simple cubic coupling which is free of derivatives
[27]. The details of the computation of radiative effects in these gauge
theories is thus considerably simplified when one uses the first order
formalism. \newline

One should, however, examine if the first and second order forms of a gauge
theory are equivalent, both at the classical level and after quantization.
Simply showing that these two forms have equivalent equations of motion may
not be adequate to establish their complete equivalence. The path integral
technique of quantization of gauge theories with a quadratic gauge fixing term
in the action (introduced by Feynman [28], deWitt [29], Mandelstam [30] and
Faddeev and Popov [31]) or with non-quadratic gauge fixing [32, 33] clearly
works for YM gauge theories and is equivalent to the path integral quantized
Yang-Mills theory obtained [34] from the canonically quantized theory
resulting from the Dirac constraint formalism [1]. However, this is contingent
upon having that part of the measure in the path integral coming from $\Delta=
\det^{1/2}\left\lbrace \theta_{a}, \theta_{b}\right\rbrace $ not contributing
to Green's functions [35, 36]. (Here $\theta_{a}$ is the set of second class
constraints in the model.) Such a contribution appears when one encounters
massive vector theories [35] or YM theories quantized on the light cone
[37-40]. However, this factor of $\Delta$ in these cases is innocuous. However
the structure of the second class constraints in the first order EH action
[18] leads to non-trivial factors of $\Delta$. Non-trivial factors of $\Delta$
also occur in a non-Abelian gauge theory involving an anti-symmetric tensor
field possessing a pseudoscalar mass [51]. This may result in there being
difficulties in establishing equivalence between the quantized form of the
first and second order forms of the EH action when using the path integral.
\newline

There have been attempts to treat in a general way the canonical structure of
first order actions [41]. However, in the general actions considered in this
reference, the possibility of tertiary (third generation) constraints arising
has not been considered, either when these new constraints are first or second
class. Hence this general analysis is deficient and needs to be extended if it
is to be applied to the first-order EH action. \newline

Attempts to treat gauge theories by using primary (first generation)
constraints to eliminate superfluous degrees of freedom [17, 25, 42] do not
distinguish between first and second class constraints nor can they be used to
find constraints beyond the first generation, and hence this approach cannot
be used in conjunction with either the approaches of refs. [4, 5] to derive
the invariances in the action present due to the occurrence of primary
constraints. In two dimensions the first and second order forms of the EH
action are no longer equivalent [14, 15]. Indeed, it has been shown in refs.
[6-9] that the first order EH action in two dimensions possesses a novel gauge
invariance that is distinct from the usual diffeomorphism invariance, but
follows from the first class constraints in the theory. It is not clear how
one could discern this new gauge invariance if the procedure of refs. [17, 25,
42] were followed in analyzing this theory. Having a knowledge of this
invariance is important if one is to quantize the first-order EH action in two
dimensions [43]. (Quantization of the second order EH in two dimensions is
non-trivial and interesting, even though the action is a total derivative [44,
45].) \newline

All of these considerations show that it is important to examine the canonical
structure of the first order form of a theory in order to determine if it is
equivalent (both before and after quantization) to its second order form.
Ultimately, we hope to address the relationship between the first and second
order form of the EH action. (Even establishing equivalence just between the
classical equations of motion arising from the first and second order form of
the EH action is non trivial; it is often ascribed to Palatini [2], though it
is in fact due to Einstein [3].) \newline

The canonical structure of the two dimensional first order EH action has
already been considered as has been noted above [6-9] with several
unanticipated features occuring. The first order EH in dimensions greater than
two is even more involved [18]; its quantization will be complicated by it
having a canonical structure which involves tertiary constraints and a
non-trivial determinant $\Delta$ arising from the second class constraints
present and so it is not clear if the Faddeev-Popov procedure followed for
first class constraints in ref. [46, 47] can be applied when quantizing the
first order form of the action. \newline

In the literature there exists some confusion about the first and second order
forms of an action. For example in ref. [10] the view is expressed that the
first and second order forms of an action are different as they have distinct
gauge invariances while in ref. [11] it is pointed out that this need not be
so as, if one uses Lagrange multipliers appropriately, one can always ensure
that the first and second order forms of an action are equivalent. \newline

We see that considering the canonical structure of first order theories that
are not as complicated as general relativity is appropriate. Maxwell
electrodynamics provides the simplest example of how the canonical formalism
can be used to establish the equivalence between the first and second order
form of a gauge theory. The canonical analysis of the second order form of
Maxwell theory appears in ref. [23]. In $2+1$ dimensions, the Maxwell
Lagrangian can be supplemented by a Chern-Simons (CS) term which results in
the vector field acquiring a mass [12]. A canonical analysis of the Abelian
and non-Abelian second order form of this massive gauge field using both
normal and light-cone coordinates appears in ref. [13]. In this paper,
canonical analysis of the first order form of these models is given and use
the first class constraints that arise is used to derive the generator of the
gauge transformation that leaves the action invariant. A canonical analysis
using light-cone coordinates is also presented. We begin by considering the
first order forms of these actions.

The standard Maxwell action for a $U(1)$ gauge field $A_{\mu}$ in three
dimensions is
\[
S_{m}=\frac{-1}{4}\int d^{3}x\left(  \partial_{\mu}A_{\nu}-\partial_{\nu
}A_{\mu}\right)  \left(  \partial^{\mu}A^{\nu}-\partial^{\nu}A^{\mu}\right)
\eqno(1)
\]
can be supplemented in three dimensions with a topological Chern-Simons
action
\[
S_{cs}=-\frac{m}{2}\int d^{3}x\epsilon^{\mu\nu\lambda}\left(  \partial_{\mu
}A_{\nu}-\partial_{\nu}A_{\mu}\right)  A_{\lambda}~,\eqno(2)
\]
so as to provide a mass to the field $A_{\mu}$ [12]. Both the actions $S_{m}$
and $S_{cs}$ are invariant under a $U(1)$ gauge transformation
\[
\delta A_{\mu}=\partial_{\mu}\zeta.\eqno(3)
\]

The action $S_{m}$ by itself can be written in first order form (i.e., the
resultant equations of motion are at most first order)
\[
S_{m}^{(1)}=\int d^{3}x\left[  \frac{1}{4}F_{\mu\nu}F^{\mu\nu}-\frac{1}%
{2}F^{\mu\nu}\left(  \partial_{\mu}A_{\nu}-\partial_{\nu}A_{\mu}\right)
\right]  ;\eqno(4)
\]
here $F_{\mu\nu}$ and $A_{\lambda}$ are independent fields [50, 26]. When the
equation of motion for $F_{\mu\nu}$ is used to eliminate $F_{\mu\nu}$ from
eq.\ (4), one recovers $S_{m}$ in eq.\ (1).

The first order form of $S_{m}+S_{cs}$ is somewhat more involved [10],
\[
S=\int d^{3}x\left[  \frac{1}{4}W^{\mu\nu}W_{\mu\nu}-\frac{1}{2}W^{\mu\nu
}\left(  \partial_{\mu}A_{\nu}-\partial_{\nu}A_{\mu}\right)  \right.
\]%
\[
\left.  -\frac{m}{2}\epsilon^{\mu\nu\lambda}W_{\mu\nu}A_{\lambda}+\frac{m^{2}%
}{2}A^{\mu}A_{\mu}\right]  .\eqno(5)
\]
Eq. (5) can be found from eqs. (1) and (2) by eliminating $F^{\mu\nu}$ from
\[
\int d^{3}x\left[  -\frac{1}{4}F_{\mu\nu}F^{\mu\nu}-\frac{m}{2}\epsilon
^{\mu\nu\lambda}F_{\mu\nu}A_{\lambda}+\frac{1}{2}W^{\mu\nu}\left(  F_{\mu\nu
}-\left(  \partial_{\mu}A_{\nu}-\partial_{\nu}A_{\mu}\right)  \right)
\right]
\]
using its equation of motion provided diag$\;g_{\mu\nu}=(+,-,-)$ and
$\epsilon_{012}=\epsilon^{012}=1$. The equations of motion for $W_{\mu\nu}$
and $A_{\mu}$ are
\[
W^{\mu\nu}=\partial^{\mu}A^{\nu}-\partial^{\nu}A^{\mu}+m\epsilon^{\mu
\nu\lambda}A_{\lambda}\eqno(6)
\]
and
\[
\partial_{\mu}W^{\mu\nu}=\frac{m}{2}\epsilon^{\alpha\beta\nu}W_{\alpha\beta
}+m^{2}A^{\nu}.\eqno(7)
\]
Substitution of $W^{\mu\nu}$ from eq.\ (6) into eq.\ (5) recovers
$S_{m}+S_{cs}$. It is apparent that the action of eq.\ (5) is invariant under
the gauge transformation of eq.\ (3) provided we also transform $W^{\mu\nu}$
\[
\delta W^{\mu\nu}=m\epsilon^{\mu\nu\alpha}\partial_{\alpha}\zeta.\eqno(8)
\]
Using eq. (7) to eliminate $A^{\nu}$ from eq. (5) results in
\[
S=\int d^{3}x\frac{-1}{2m^{2}}\left(  \partial_{\alpha}W^{\alpha\mu}%
\partial^{\beta}W_{\beta\mu}+m\epsilon_{\alpha\beta\gamma}W^{\mu\gamma
}\partial_{\mu}W^{\alpha\beta}\right)  ,
\]
which is an alternate second order way of re-expressing $S_{cs}+S_{m}$.
Further discussion of how to rewrite the action for Abelian gauge fields when
topological actions also occur appear in refs. [19,20]. An action for the
Chern-Simons model when accompanied by a Stueckelberg mass term is in ref. [21].

We now will demonstrate how the Dirac analysis of constrained systems [1] can
be used to analyze actions which are first order in derivatives by applying
this procedure to the actions of eqs.\ (4) and (5). It is of particular
interest to show how this approach can be used to derive the gauge invariances
of eqs.\ (3) and (8). It is also possible to generate a non-Abelian version of
eqs.\ (5 - 8) which we provide in section 5. The canonical structure of these
models when expressed in light-cone coordinates is examined in section 6.

\section{Maxwell Electrodynamics}

The action of eq.\ (4) in four dimensional space (with signature (+ + + $-$))
can be written as
\[
S_{m}^{(1)}=\int d^{4}x\left[  \frac{1}{2}\left(  \vec{B}^{2}-\vec{E}%
^{2}\right)  +\vec{E}\cdot\left(  \dot{\vec{A}}+\nabla A\right)  -\vec{B}%
\cdot\nabla\times\vec{A}\right]  \eqno(9)
\]
where
\[
B^{i}=\frac{1}{2}\epsilon^{ijk}F^{jk},\;\;\;\;E^{i}=F^{i0},\;\;\;\;A=A^{0}%
.\eqno(10)
\]
The momenta conjugate to $\vec{B}$, $\vec{E}$, $\vec{A}$ and $A$ give rise to
ten primary constraints [9]
\[
\vec{\Pi}_{B}=0,\;\;\;\;\vec{\Pi}_{E}=0,\;\;\;\;\vec{\pi}=\vec{E}%
,\;\;\;\;\pi=0,\eqno(11-14)
\]
with eqs.\ (12, 13) immediately been seen to be a pair of second class
constraints. Which class the constraints of eqs. (11, 14) belong to can only
be decided upon after the secondary constraints have been determined.

The canonical Hamiltonian associated with eq.\ (9) is
\[
H_{c} = \int\left[ \frac{1}{2} \left( \vec{E}^{2} - \vec{B}^{2}\right)  -
\vec{E} \cdot\nabla A + \vec{B} \cdot\nabla\times\vec{A}\right] d^{4}x
.\eqno(15)
\]
This is supplemented with Lagrange multiplier fields that ensure that the
constraints are satisfied to yield the total Hamiltonian
\[
H_{T} = \int\left[ H_{c} + \vec{\Lambda}_{B} \cdot\vec{\Pi}_{B} + \vec
{\Lambda}_{E} \cdot\vec{\Pi}_{E} + \vec{\lambda} \cdot\left( \vec{\pi} -
\vec{E}\right)  + \lambda\pi\right] d^{4}x .\eqno(16)
\]
The Lagrangian equations of motion are equivalent to the Hamiltonian equations
of motion in which the total Hamiltonian has been used [22].

For consistency, the constraints must be time independent and so must have a
vanishing Poisson bracket (PB) with $H_{T}$. We hence find that [9]
\[
\left\{  \vec{\Pi}_{B},H_{T}\right\}  =\vec{B}-\nabla\times\vec{A},\eqno(17)
\]%
\[
\left\{  \vec{\Pi}_{E},H_{T}\right\}  =-\vec{E}+\nabla A+\vec{\lambda
},\eqno(18)
\]%
\[
\left\{  \vec{\pi}-\vec{E},H_{T}\right\}  =-\nabla\times\vec{B}-\vec{\Lambda
}_{E},\eqno(19)
\]%
\[
\left\{  \pi,H_{T}\right\}  =-\nabla\cdot\vec{E}\eqno(20)
\]
must vanish.

By eq. (17), we see that there is now an additional secondary constraint
$\vec{B} - \nabla\times\vec{A} = 0$; it is immediately apparent that now both
this constraint and that of eq. (11) are second class. Furthermore, from eq.
(20) we see that the longitudinal component of $\vec{E}$ must vanish; keeping
mind eq. (12) this is seen to be a second class constraint. This and eq. (13)
result in $\nabla\cdot\vec{\pi} = 0$ being a secondary, first class constraint.

The time derivatives of $\vec{B}-\nabla\times\vec{A}$ and $\nabla\cdot\vec{E}$
lead to
\[
\left\{  \vec{B}-\nabla\times\vec{A},H_{T}\right\}  =\vec{\Lambda}_{B}%
-\nabla\times\vec{\lambda}\eqno(21)
\]
and
\[
\left\{  \nabla\cdot\vec{E},H_{T}\right\}  =\nabla\cdot\vec{\Lambda}%
_{E}.\eqno(22)
\]
We now see that the time derivative of the secondary constraints fix the
Lagrange multipliers $\vec{\lambda}$ (from eq. (18)), the transverse part of
$\vec{\Lambda}_{E}$ (from eq. (19)), $\vec{\Lambda}_{B}$ (from eq. (21)) and
the longitudinal part of $\vec{\Lambda}_{E}$ (from eq. (22)). The only
undetermined Lagrange multiplier is $\lambda$ and the only first class
constraints are the primary constraint $\pi=0$ and the secondary constraint
$\nabla\cdot\vec{\pi}=0$. As expected [1], the number of arbitrary functions
is one (the Lagrange multiplier $\lambda$); this equals the number of primary
first class constraints $(\pi=0)$ and the number of gauge parameters ($\zeta$
in eq. (3)). We also note that we have twelve second class constraints and two
first class constraints, which when combined with two gauge conditions, gives
sixteen constraints in total. There are twenty degrees of freedom ($F_{\mu\nu
}$, $A_{\lambda}$ and their conjugate momenta) initially, and hence the number
of physical degrees of freedom is $20-16=4$ (the two transverse polarizations
and their conjugage momenta).

When eliminating the second class constraints ($\vec{\Pi}_{B} = \vec{\Pi}_{E}
= \vec{\pi} -\vec{E} = \vec{B} - \nabla\times\vec{A} = 0$) through the
introduction of Dirac Brackets [1], which replace the Poisson Brackets, one
finds that the Dirac Brackets are identical to the Poisson Brackets except
that now
\[
\left\{  B_{i}\left(  \vec{x},t\right)  ,E _{j}\left(  \vec{y},t\right)
\right\}  ^{\ast}= \epsilon_{ipj} \partial_{p}^{x} \delta\left(  \vec{x}%
-\vec{y}\right) . \eqno(23)
\]
This Dirac Bracket is identical to what would be obtained if the Dirac
constraint procedure were applied to the second order Maxwell action of eq.
[1] and the Coulomb gauge condition $\nabla\cdot\vec{A} = 0$ were employed in
conjunction with the Gauss law constraint, which is first class [23]. The
Dirac Bracket of eq. (23) has, in contrast, been derived by elimination of the
second class constraints arising from the first order Maxwell action of eq.
[4] without imposing any gauge condition. The form of $H_{T} $ is, upon
elimination of the second class constraints
\[
H_{T}=\frac{1}{2}\left(  \vec{\pi}^{2}+(\nabla\times\vec{A})^{2}\right)
+A\nabla\cdot\vec{\pi}+\lambda\pi.\eqno(24)
\]
The field $A$ becomes a Lagrange multiplier field. Eq.\ (24) gives the same
expression for $H_{T}$ that one obtains if the second order form for $S_{m}$
in eq.\ (1) is treated using the Dirac procedure [1].

An approach to determine the gauge transformation that leaves the action
invariant is in ref. [4] while an approach based on the equations of motion is
in ref. [5]. In both cases we find that the gauge generator is
\[
G(\zeta,\dot{\zeta})=-\int d^{3}x\left(  \zeta\nabla\cdot\vec{\pi}+\dot{\zeta
}\pi\right) , \eqno(25)
\]
so that
\[
\delta A=\left\{  A,G\right\}  ^{\ast}=-\dot{\zeta}\eqno(26)
\]
and
\[
\delta\vec{A}=\left\{  \vec{A},G\right\}  ^{\ast}=\nabla\zeta,\eqno(27)
\]
while
\[
\delta\vec{E}=\left\{  \vec{E},G\right\}  ^{\ast}=0=\delta\vec{B}.\eqno(28)
\]
Together then, $\delta A_{\mu}=\partial_{\mu}\zeta$ and $\delta F_{\mu\nu}=0$,
as one would expect from inspection of $S_{m}^{(1)}$ in eq.~(4).

We now apply the Dirac formalism to the more interesting (and complicated)
case of the action $S$ of eq.\ (5).

\section{Topologically Massive Electrodynamics}

The action of eq.\ (5) can be written as
\[
S=\int d^{3}x\left[  \frac{1}{2}\left(  W^{2}-\vec{W}^{2}\right)  +\frac
{m^{2}}{2}\left(  A^{2}-\vec{A}^{2}\right)  -\left(  W\nabla\times\vec{A}%
+\vec{W}\cdot\dot{\vec{A}}+\vec{W}\cdot\nabla A\right)  \right.
\]%
\[
\left.  -m\left(  \vec{W}\times\vec{A}+WA\right)  \right]  \eqno(29)
\]
if the metric is diagonal $(++-)$, $\epsilon_{012}=1$, $A=A^{0}$, $W=\frac
{1}{2}\epsilon_{ij}W^{ij}$, $W^{i}=W^{0i}$ and $\vec{U}\times\vec{V}%
=\epsilon_{ij}U^{i}V^{j}$. The momenta associated with $A$, $\vec{A}$, $W$ and
$\vec{W}$ are now given by the primary constraints
\[
\pi=0,\;\;\;\vec{\pi}+\vec{W}=0,\;\;\;\Pi=0,\;\;\;\vec{\Pi}=0,\eqno(30-33)
\]
respectively. The constraints of eqs.\ (31, 33) are second class; if DBs are
used it is possible to immediately replace $\vec{W}$ by $-\vec{\pi}$ in the
canonical Hamiltonian and we obtain
\[
H_{c}=\frac{1}{2}\left(  \vec{\pi}^{2}-W^{2}\right)  +\frac{m^{2}}{2}\left(
\vec{A}^{2}-A^{2}\right)  +W\nabla\times\vec{A}+A\nabla\cdot\vec{\pi}+m\left(
AW+\vec{\pi}\times\vec{A}\right)  .\eqno(34)
\]

Consistency means that $\dot{\Pi} = \left\lbrace \Pi, H_{c} \right\rbrace $
should weakly vanish; with $H_{c}$ given by eq.\ (34) then
\[
\left\lbrace \Pi, H_{c}\right\rbrace = W - \nabla\times\vec{A} - m A = 0
\eqno(35)
\]
is a secondary constraint. Similarly, as $\dot{\pi} = 0$, we obtain another
secondary constraint,
\[
\left\lbrace \pi, H_{c}\right\rbrace = m^{2} A - \nabla\cdot\vec{\pi} - mW =
0. \eqno(36)
\]

Eqs.\ (30, 32, 35, 36) together form four constraints. However, the PBs of
these four constraints form a matrix with rank two; consequently appropriate
linear combinations of these four constraints can be chosen so that two are
first class and two are second class. A suitable pair of first class
constraints are
\[
\gamma_{1}=\pi+m\Pi,\;\;\;\;\;\;\;\gamma_{2}=m\nabla\times\vec{A}+\nabla
\cdot\vec{\pi}\eqno(37,38)
\]
and of second class constraints are (provided $m\neq0$)
\[
\chi_{1}=\Pi,\;\;\;\;\;\;\;\chi_{2}=\nabla\times\vec{A}+mA-W.\eqno(39,40)
\]
Using eq.\ (40) to eliminate $W$ in eq.\ (34) leads to
\[
H_{c}=\frac{1}{2}\left[  \vec{\pi}^{2}+\left(  \nabla\times\vec{A}\right)
^{2}+m^{2}\vec{A}^{2}\right]  +A\left(  m\nabla\times\vec{A}+\nabla\cdot
\vec{\pi}\right)  +m\vec{\pi}\times\vec{A}.\eqno(41)
\]
It is now evident that
\[
\left\{  \gamma_{1},H_{c}\right\}  =-\gamma_{2},\;\;\;\;\left\{  \gamma
_{2},H_{c}\right\}  =0,\;\;\;\;\left\{  \gamma_{1},\gamma_{2}\right\}
=0.\eqno(42-44)
\]
With the first class constraints of eqs.\ (37, 38) satisfying the commutation
relations of eqs.\ (42-44), the methods of refs. [5, 6] lead to the gauge
generator
\[
G=\int d^{2}x\left[  -\dot{\zeta}\pi+\zeta(m\nabla\times\vec{A}+\nabla
\cdot\vec{\pi})\right]  ,\eqno(45)
\]
so that
\[
\delta A=\left\{  A,G\right\}  =-\dot{\zeta}\eqno(46)
\]
and
\[
\delta\vec{A}=\left\{  \vec{A},G\right\}  =\nabla\zeta,\eqno(47)
\]
as well as
\[
\delta\pi_{i}=m\epsilon_{ij}\partial_{j}\zeta.\eqno(48)
\]
Eqs.\ (31) and (40) can now be used to show that
\[
\delta W=-m\dot{\zeta},\eqno(49)
\]%
\[
\delta W_{i}=-m\epsilon_{ij}\partial_{j}\zeta.\eqno(50)
\]
Together, from eqs.\ (46, 47, 49, 50) we recover the gauge transformations of
eqs.\ (3, 8).

\section{Non-Abelian Model}

The canonical structure of the first order form of the Maxwell and Maxwell
plus Chern-Simons actions have been analyzed in some detail above. In fact, a
first order form of the non-Abelian extension of these models can also be
considered. A canonical analysis of the second order form of these
topologically massive gauge theories is carried out in ref.\ [13] using normal
and light-cone coordinates. Here we present a canonical analysis of the first
order form without imposition of any particular coordinate system. We start
with the Lagrangian [12]%

\[
L=-\frac{1}{4}\left(  \partial_{\mu}A_{v}^{a}-\partial_{v}A_{\mu}^{a}%
+f^{abc}A_{\mu}^{b}A_{\nu}^{c}\right)  \left(  \partial^{\mu}A^{av}%
-\partial^{v}A^{a\mu}+f^{abc}A^{b\mu}A^{cv}\right)
\]
\nonumber%
\[
-m\epsilon^{\mu v\lambda}\left(  \partial_{\mu}A_{v}^{a}A_{\lambda}^{a}%
+\frac{1}{3}f^{abc}A_{\mu}^{a}A_{v}^{b}A_{\lambda}^{c}\right)  \eqno(51)
\]
which is invariant under the transformation%

\[
\delta A^{a}_{\mu}= \partial_{\mu}\theta^{a} + f^{abc} A^{b}_{\mu}\theta^{c}
\equiv D^{ab}_{\mu}\theta^{b} .\eqno(52)
\]
The canonical structure of this action was considered in ref. [13].

The Lagrangian of eq.\ (51) can be derived from%

\[
L = - \frac{1}{4} F^{a}_{\mu v} F^{a \mu v} - \frac{m}{2} \epsilon^{\mu v
\lambda} \left( F^{a}_{\mu v} A^{a}_{\lambda}- \frac{1}{3} f^{abc} A^{a}_{\mu
}A^{b}_{v} A^{c}_{\lambda}\right)
\]
\nonumber
\[
+ \frac{1}{2} W^{a \mu v} \left[ F^{a}_{\mu v} - \left( \partial_{\mu}%
A^{a}_{v} - \partial_{v} A^{a}_{\mu}+ f^{abc} A^{b}_{\mu}A^{c}_{v} \right)
\right]  .\eqno(53)
\]

Upon making use of the equation of motion for the Lagrange multiplier field
$W^{a}_{\mu v}$, the field strength $F^{a}_{\mu v}$ undergoes the
transformation
\[
\delta F^{a}_{\mu v} = f^{abc} F^{b}_{\mu v} \theta^{c}, \eqno(54)
\]

\noindent when $A_{\mu}^{a}$ transforms according to eq.\ (52). The equation
of motion for $F_{\mu v}^{a}$ however shows that
\[
F_{\mu v}^{a}=W_{\mu v}^{a}-m\epsilon_{\mu v\lambda}A^{a\lambda},\eqno(55)
\]
which can be used to eliminate $F_{\mu v}^{a}$ in eq.\ (53), yielding
\[
L=\frac{1}{4}W_{\mu v}^{a}W^{a\mu v}-\frac{1}{2}W^{a\mu v}\left(
\partial_{\mu}A_{v}^{a}-\partial_{v}A_{\mu}^{a}+f^{abc}A_{\mu}^{b}A_{v}%
^{c}\right)
\]%
\[
+\frac{m^{2}}{2}A_{\mu}^{a}A^{a\mu}-\frac{m}{2}\epsilon^{\mu v\lambda}\left(
W_{\mu v}^{a}A_{\lambda}^{a}-\frac{1}{3}f^{abc}A_{\mu}^{a}A_{v}^{b}A_{\lambda
}^{c}\right)  \eqno(56)
\]

\noindent Furthermore, together, eqs.\ (52, 54, 55) show that the
transformation of $W^{a}_{\mu v}$ takes the form%

\[
\delta W_{\mu v}^{a}=m\epsilon_{\mu v\lambda}\partial^{\lambda}\theta
^{a}+f^{abc}W_{\mu v}^{b}\theta^{c}.\eqno(57)
\]

This transformation when combined with eq.\ (52) leaves eq.\ (56) invariant.
Eq.\ (56) is a non-Abelian modification of eq.\ (5) and can also be analyzed
using Dirac's procedure.

As with the Abelian model in the preceding section, we define%

\[
A^{a} = A^{a0}\;, \hspace{1cm} W^{a} = \frac{1}{2} \epsilon_{ij} W^{a}_{ij}\;,
\hspace{1cm} W^{ai} = W^{a0i} \;,
\]
so that%

\[
L = - \vec{W}^{a} \cdot{\dot{\vec{A}}}^{a} + \frac{1}{2} (W^{a} W^{a} -
\vec{W}^{a} \cdot\vec{W}^{a}) + \frac{m^{2}}{2} (A^{a} A^{a} - \vec{A}^{a}
\cdot\vec{A}^{a})
\]
\[
+ A^{a} (\nabla\cdot\vec{W}^{a} + f^{abc} \vec{A}^{b} \cdot\vec{W}^{c}) -
W^{a} (\nabla\times\vec{A}^{a} + \frac{1}{2} f^{abc} \vec{A}^{b} \times\vec
{A}^{c})
\]
\[
-m (W^{a} A^{a} + \vec{W}^{a} \times\vec{A}^{a}) + \frac{m}{2} A^{a} f^{abc}
\vec{A}^{b} \times\vec{A}^{c}. \eqno(58)
\]

The momenta conjugate to $A^{a}, A^{a}_{i}, W^{a}$ and $W^{a}_{i}$ result in
the constraint equations%

\[
\pi^{a} = 0, \hspace{1cm} \vec{\pi}^{a} + \vec{W}^{a} = 0, \hspace{1cm}
\Pi^{a} = 0, \hspace{1cm} \vec{\Pi}^{a} = 0, \eqno(59-62)
\]

\noindent much like eqs.\ (30-33). Eqs.\ (60, 61) are second class constraints
and can be used to replace $\vec{W}^{a}$ by $-\vec{\pi}^{a}$. Eqs.\ (59, 62)
are primary constraints which imply the secondary constraints%

\[
-m^{2} A^{a} + ( \nabla\cdot\vec{\pi}^{a} + f^{abc} \vec{A}^{b} \cdot\vec{\pi
}^{c}) + m W^{a} - \frac{m}{2} f^{abc} \vec{A}^{b} \times\vec{A}^{c} = 0
\eqno(63)
\]
and%

\[
-W^{a} + (\nabla\times\vec{A}^{a} + \frac{1}{2} f^{abc} \vec{A}^{b} \times
\vec{A}^{c}) + m A^{a} = 0 \eqno(64)
\]
respectively. Of the four constraints of eqs.\ (59, 62, 63, 64), two linear
combinations can be taken to be first class%

\[
\gamma_{1}=\pi^{a}+m\Pi^{a},\hspace{2cm}\gamma^{2}=\nabla\cdot\vec{\pi}%
^{a}+f^{abc}\vec{A}^{b}\cdot\vec{\pi}^{c}+m\nabla\times\vec{A}^{a}%
\eqno(65,66)
\]
and two to be second class
\[
\chi_{1}=\Pi^{a},\hspace{2cm}\chi_{2}=-W^{a}+(\nabla\times\vec{A}^{a}+\frac
{1}{2}f^{abc}\vec{A}^{b}\times\vec{A}^{c})+mA^{a}.\eqno(67,68)
\]
with no further constraints being required. Eqs.\ (65-68) are generalizations
of eqs.\ (37-40). The first class constraints can now be used to generate the
transformation of eqs.\ (52, 57).

It is of interest to define
\[
X_{\mu}^{a}=\frac{1}{2}\left(  A_{\mu}^{a}-\frac{1}{2m}\epsilon_{\mu
\lambda\sigma}W^{a\lambda\sigma}\right)  ,\eqno(69a)
\]%
\[
Y_{\mu}^{a}=\frac{1}{2}\left(  A_{\mu}^{a}+\frac{1}{2m}\epsilon_{\mu
\lambda\sigma}W^{a\lambda\sigma}\right)  ,\eqno(69b)
\]
so that the Lagrangian of eq. (56) can be rewritten as
\[
\mathcal{L}=-m\epsilon^{\mu\nu\lambda}\left(  X_{\mu}^{a}\partial_{\nu
}X_{\lambda}^{a}+\frac{1}{3}f^{abc}X_{\mu}^{a}X_{\nu}^{b}X_{\lambda}%
^{c}\right)  -2m^{2}Y_{\mu}^{a}Y^{a\mu}%
\]

\[
+m\epsilon^{\mu\nu\lambda}\left[  Y_{\mu}^{a}\partial_{\nu}Y_{\lambda}%
^{a}+f^{abc}\left(  Y_{\mu}^{a}Y_{\nu}^{b}X_{\lambda}^{c}+\frac{2}{3}Y_{\mu
}^{a}Y_{\nu}^{b}Y_{\lambda}^{c}\right)  \right]  .\eqno(70)
\]
\qquad The terms in eq. (70) that depend solely on $X_{\mu}^{a}$ are pure
Chern-Simons. From eqs. (52, 57, 69), eq. (70) is invariant under the
transformations
\[
\delta X_{\mu}^{a}=\partial_{\mu}\theta^{a}+f^{abc}X_{\mu}^{b}\theta
^{c},\eqno(71a)
\]%
\[
\delta Y_{\mu}^{a}=f^{abc}Y_{\mu}^{b}\theta^{c}.\eqno(71b)
\]

\section{Canonical Analysis using Light-Cone Coordinates}

Light-cone coordinates were originally introduced by Dirac [48] and have been
used in a variety of circumstances [37-40, 50]. The coordinates in an
$N$-dimensional space are taken to be%

\[
x^{\pm}=\frac{(x^{0}\pm x^{N-1})}{\sqrt{2}}\;\;,\quad x^{i}=(x^{1}%
,x^{2},\ldots,x^{N-2});\eqno(72a,b)
\]
so that if $A^{\mu\nu}$ and $B^{\mu\nu}$ are antisymmetric and $N=3$
\[
a\cdot b=a^{+}b^{-}+a^{-}b^{+}-a^{i}b^{i},\;\;A^{\mu\nu}B_{\mu\nu}%
=-2A^{+-}B^{+-}-2A^{+i}B^{-i}-2A^{-i}B^{+i}+A^{ij}B^{ij}\;,\eqno(73a,b)
\]%
\[
\epsilon_{\mu\nu\lambda}A^{\mu}B^{\nu}C^{\lambda}=A^{1}(B^{+}C^{-}-B^{-}%
C^{+})+\left(  \text{cyc. perm.}\right)  .\eqno(73c)
\]
Since $\partial^{2}=\partial^{+}\partial^{-}-\partial^{i}\partial^{i}$, any
canonical analysis using $x^{+}$ as the \textquotedblleft
time\textquotedblright\ variable will lead to having a first-order action
(ie., one that is first order in $\partial^{+}$).

For example, the Lagrangian of eq. (1) when using light-cone coordinates
becomes
\[
{\mathcal{L}}=\frac{1}{2}f^{+-}f^{+-}+f^{+i}f^{-i}-\frac{1}{4}f^{ij}%
f^{ij}\eqno(74)
\]
where $f^{\mu\nu}=\partial^{\mu}A^{\nu}-\partial^{\nu}A^{\mu}$. The canonical
momenta conjugate to $A^{+}$, $A^{-}$ and $A^{i}$ are
\[
\pi_{+}=0\;,\qquad\pi_{-}=f^{+-}\;,\qquad\pi_{i}=f^{-i}.\eqno(75a,b,c)
\]
Eq. (75c) is a primary second class constraint while the Hamiltonian
\[
{\mathcal{H}}=\pi_{+}\partial^{+}A^{+}+\pi_{-}\partial^{+}A^{-}+\pi
_{i}\partial^{+}A^{i}-{\mathcal{L}}\eqno(76)
\]%
\[
=\frac{1}{2}\pi_{-}^{2}+\frac{1}{4}f^{ij}f^{ij}-A^{+}(\partial^{-}\pi
_{-}+\partial^{i}\pi_{i})
\]
and the constraint of eq. (75a) leads to the secondary constraint
\[
(\partial^{-}\pi_{-}+\partial^{i}\pi_{i})=0.\eqno(77)
\]
Both eqs.(75a) and (77) are first class constraints; there are no tertiary
constraints. With $2N$ variables in phase space, $N-2$ second class
constraints, two first class constraints and two associated gauge conditions
(eg. $A^{+}=A^{-}=0$ or $A^{+}=\partial^{i}A^{i}=0$) there are
$2N-(N-2)-2-2=N-2$ degrees of freedom. Using $x^{0}$ as the time variable
leads to $2(N-2)$ degrees of freedom as there are no second class constraints
in this case.

If we consider the Lagrangian in eq. (4) then this becomes using light-cone
coodinates
\[
{\mathcal{L}}=-\frac{1}{2}F^{+-}F^{+-}-F^{+i}F^{-i}+\frac{1}{4}F^{ij}%
F^{ij}\eqno(78)
\]%
\[
+F^{+-}f^{+-}+F^{+i}f^{-i}+F^{-i}f^{+i}-\frac{1}{2}F^{ij}f^{ij}.
\]
There are now the primary second class constraints
\[
\Pi_{+-}=0=\Pi_{-i}=\pi_{-}-F^{+-}=\pi_{i}-F^{-i}=\Pi_{ij}\;,\eqno(79a-e)
\]
the secondary second class constraints
\[
F^{-i}-f^{-i}=0,\eqno(80)
\]
the primary first class constraints
\[
\pi_{+}=0=\Pi_{+i}\eqno(81a,b)
\]
and the secondary first class constraint
\[
\partial^{-}\pi_{-}+\partial^{i}\pi_{i}=0.\eqno(82)
\]
Here $\Pi_{+-}$, $\Pi_{+i}$ $\Pi_{-i}$ are the momenta conjugate to $F^{+-}$,
$F^{+i}$ and $F^{-i}$. There are $\frac{N(N+1)}{2}$ fields in the initial
Lagrangian ($F_{\mu\nu}$ and $A_{\mu}$); once all of the constraints are taken
into account there are $(N-2)$ independent fields in phase space as before. In
the gauge in which $A^{+}=0$ (paired with the consraint of eq. (81a)) the
Hamiltonian reduces to
\[
{\mathcal{H}}=\frac{1}{2}\left(  -\frac{1}{\partial^{-}}\partial^{i}\pi
_{i}\right)  ^{2}+\frac{1}{4}f^{ij}f^{ij}\eqno(83)
\]
with the Dirac Brackets
\[
\left\{  A^{i}(x),A_{j}(y)\right\}  ^{\ast}=\frac{1}{2\partial^{-}}%
\delta(x-y)\delta_{j}^{i}\;,\eqno(84a)
\]%
\[
\left\{  A^{i}(x),\pi_{j}(y)\right\}  ^{\ast}=\frac{1}{2}\delta(x-y)\delta
_{j}^{i}.\eqno(84b)
\]
If in conjunction with the constraint of eq. (82) we choose the gauge
condition $\partial^{i}A_{i}=0$, then $\delta_{j}^{i}$ in eq. (84) gets
replaced by $\delta_{j}^{i}-\partial^{i}\partial_{j}/\partial^{2}$, which in
three dimensions equals zero.

We now examine the action of eq. (56) in light-cone coordinates. The
Lagrangian now becomes
\begin{align*}
{\mathcal{L}}  & =-\frac{1}{2}W^{a+-}W^{a+-}-W^{a+1}W^{a-1}+m\left(
W^{a1-}A^{a+}+W^{a+1}A^{a-}+W^{a-+}A^{a1}\right)  \\
& +m^{2}\left(  A^{a+}A^{a-}-\frac{1}{2}A^{a1}A^{a1}\right)
\end{align*}%
\[
+\left(  W^{a+-}f^{a+-}+W^{a+1}f^{a-1}+W^{a-1}f^{a+1}\right)  +m\epsilon
^{abc}(A^{a1}A^{b+}A^{c-})\eqno(85)
\]
where $f^{a\mu\nu}\equiv\partial^{\mu}A^{a\nu}-\partial^{\nu}A^{a\mu}%
+\epsilon^{abc}A^{b\mu}A^{c\nu}$.

With this Lagrangian, the primary second class constraints are
\[
\Pi_{+-}^{a}=\Pi_{-1}^{a}=\pi_{1}^{a}-W^{a-1}=\pi_{-}^{a}-W^{a+-}%
=0\eqno(86a-d)
\]
where $\Pi_{\mu\nu}^{a}$ is the momentum conjugate to $W^{a\mu\nu}$ and
$\pi_{\mu}^{a}$ is the momentum conjugate to $A^{a\mu}$. The primary first
class constraints are
\[
\pi_{+}^{a}=\Pi_{+1}^{a}=0.\eqno(87a,b)
\]
The Hamiltonian thus can be written as
\[
\hspace{-4.5cm}{\mathcal{H}}=\frac{1}{2}(\pi_{-}^{a})^{2}+W^{a+1}\left(
\pi_{1}^{a}-mA^{a-}-f^{a-1}\right)  +m\pi_{-}^{a}A^{a1}+\frac{1}{2}%
m^{2}(A^{a1})^{2}%
\]%
\[
+A^{a+}\left[  -(D^{1}\pi_{1})^{a}-(D^{-}\pi_{-})^{a}+m\pi_{1}^{a}-m^{2}%
A^{a-}-m\epsilon^{abc}A^{b-}A^{c1}\right]  \left(  D_{\mu}^{ab}\equiv
\partial_{\mu}\delta^{ab}+\epsilon^{apb}A_{\mu}^{\rho}\right)  ,\eqno(88)
\]
and so we find the secondary constraints
\[
\pi_{1}^{a}-mA^{a-}-f^{a-1}=0\;,\eqno(89a)
\]%
\[
(D^{1}\pi_{1})^{a}+(D^{-}\pi_{-})^{a}-m\left(  \pi_{1}^{a}-mA^{a-}%
-\epsilon^{abc}A^{b-}A^{c1}\right)  =0.\eqno(89b)
\]
These constraints are second and first class respectively. The constraint of
eq. (89b), upon using eq. (89a), becomes
\[
(D^{1}\pi_{1})^{a}+(D^{-}\pi_{-})^{a}+m\left(  \partial^{1}A^{a-}-\partial
^{-}A^{a1}\right)  =0.\eqno(90)
\]
The use of a second class constraint in this way must be accompanied by
replacement of Poisson Brackets by Dirac Brackets [1]. One peculiarity of this
system is that there are naively five second class constraints; one normally
anticipates an even number of second class constraints. (Only one dynamical
degree of freedom is physical.) Having an odd number of constraints is not a
problem though as the Poisson Bracket of the constraints of eq. (89a)
\[
\left\{  \pi_{1}^{a}(x)-mA^{a-}(x)-f^{a-1}(x),\;\;\;\pi_{1}^{b}(y)-mA^{b-}%
(y)-f^{b-1}(y)\right\}
\]%
\[
=-2D_{x}^{ab-}\delta(x-y)\eqno(91)
\]
is non-local. The Dirac Brackets which differ from Poisson Brackets following
from eq. (91) are
\[
\left\{  A^{a1}(x),\pi_{1}^{b}(y)\right\}  ^{\ast}=\frac{1}{2}\delta
^{ab}\delta(x-y)\;,\eqno(92a)
\]%
\[
\left\{  A^{a1}(x),A^{b1}(y)\right\}  ^{\ast}=\frac{-1}{2D_{x}^{ab-}}%
\delta(x-y)\;,\eqno(92b)
\]%
\[
\left\{  A^{a1}(x),\pi_{-}^{b}\right\}  ^{\ast}=\frac{-1}{2D_{x}^{ac-}}\left(
m\delta^{cb}-D_{x}^{cb1}\right)  \delta(x-y)\;,\eqno(92c)
\]%
\[
\left\{  \pi_{1}^{a}(x),\pi_{1}^{b}(y)\right\}  ^{\ast}=\frac{1}{2}D_{x}%
^{ab-}\delta(x-y)\;,\eqno(92d)
\]%
\[
\left\{  \pi_{1}^{a}(x),\pi_{-}^{b}(y)\right\}  ^{\ast}=\frac{1}{2}\left(
m\delta^{ab}-D_{x}^{ab1}\right)  \delta(x-y)\;,\eqno(92e)
\]%
\[
\left\{  \pi_{-}^{a}(x),\pi_{-}^{b}(y)\right\}  ^{\ast}=-\left[  (m+D_{x}%
^{1})\left(  \frac{1}{2D_{x}^{-}}\right)  (m-D_{x}^{1})\right]  ^{ab}%
\delta(x-y).\eqno(92f)
\]
If we accompany the primary first class constraint of eq. (87a) with the gauge
condition $A^{a+}=0$, then by use of the second class constraints, the
Hamiltonian of eq. (88) reduces to the simple form
\[
{\mathcal{H}}=\frac{1}{2}\left(  \pi_{-}^{a}+mA^{a1}\right)  ^{2},\eqno(93)
\]
with none of the Dirac Brackets of eq. (92) being affected. Neither eqs. (92)
nor (93) are changed if we adopt the gauge condition $W^{a+1}=0$ in
conjunction with the first class constraint of eq. (87b).

It is now possible to adopt the gauge condition
\[
\pi_{-}^{a} = 0\eqno(94)
\]
in conjunction with the constraint of eq. (90) provided $m \neq0$. This
reduces the Hamiltonian of eq. (93) to the simple form
\[
{\mathcal{H}} = \frac{1}{2} m^{2} (A^{a1})^{2}.\eqno(94)
\]
The appropriate Dirac Bracket serves to eliminate the constraints $\theta_{a}$
of eqs. (89a, 90, 94). This involves inverting the matrix $\Delta_{ab}
\equiv\left\lbrace \theta_{a}, \theta_{b} \right\rbrace $ which is given by
\[
\Delta_{ab} = \left(
\begin{array}
[c]{lll}%
-2D^{ab-} & \epsilon^{abc}\left( \pi_{1}^{c}-f^{c-1}-mA^{c-}\right)  & D^{ab1}
-m \delta^{ab}\\
&  & \\
\epsilon^{abc}\left( \pi_{1}^{c}-f^{c-1}-mA^{c-}\right)  & \epsilon
^{abc}\left[  (D^{-}\pi_{-})^{c} + (D^{1}\pi_{1})^{c}\right.  & \epsilon
^{abc}\pi_{1}^{c}-m \delta^{ab}\partial^{1}\\
& \left.  +m (\partial^{-}A^{c1}-\partial^{1}A^{c-})\right]  & D^{ab1} +
m\delta^{ab}\\
&  & \\
D^{ab1} + m\delta^{ab} & \epsilon^{abc}\pi_{-}^{c} -m\delta^{ab}\partial^{1} &
\quad0
\end{array}
\right) . \eqno(95)
\]
Since once the Dirac Brackets are used in place of Poisson Brackets, the
constraints $\theta_{a}$ can simply be set equal to zero, it is sufficient to
use the inverse of
\[
\Delta_{ab}^{(0)} = \left(
\begin{array}
[c]{ccc}%
-2D^{ab-} & 0 & D^{ab1}-m\delta^{ab}\\
&  & \\
0 & 0 & -m\delta^{ab}\partial^{1}\\
&  & \\
D^{ab1} + mf^{ab} & -m\delta^{ab}\partial^{1} & 0
\end{array}
\right)  \eqno(96)
\]
when defining the Dirac Bracket. Since
\[
\Delta_{ab}^{(0)-1} = \left(
\begin{array}
[c]{ccc}%
\frac{-1}{2D^{-}} & \frac{-1}{2m}\frac{1}{D^{-}}(D^{1}-m)\frac{1}{\partial
^{1}} & 0\\
&  & \\
\frac{-1}{2m}\frac{1}{\partial^{1}}(D^{1}+m)\frac{1}{D^{-}} & \frac{-1}%
{2m^{2}}\frac{1}{\partial^{1}}(D^{1}+m)\frac{1}{D^{-}}(D^{1}-m)\frac
{1}{\partial^{1}} & \frac{-1}{m\partial^{1}}\\
&  & \\
0 & \frac{-1}{m\partial^{1}} & 0
\end{array}
\right) ^{ab} \eqno(97)
\]
the Dirac Bracket
\[
\left\lbrace M,N\right\rbrace ^{\ast} = \left\lbrace M,N\right\rbrace -
\left\lbrace M,\theta_{a}\right\rbrace \Delta_{ab}^{(0)-1} \left\lbrace
\theta_{b},N\right\rbrace \eqno(98)
\]
for the single dynamical degree of freedom $A^{a1}$ in phase space reduces to
\[
\left\lbrace A^{a1}(x^{1}, x^{-}, t^{+}), \qquad A^{b1}(y^{1}, y^{-}, t^{+})
\right\rbrace ^{\ast}
\]
\[
= \frac{-1}{2} \left[  \frac{1}{D^{-}} + \frac{1}{m} D^{1} \frac{1}%
{\partial^{1}} (D^{1}+m) \frac{1}{D^{-}} + \frac{1}{m}\frac{1}{D^{-}}
(D^{1}-m)\frac{1}{\partial^{1}}D^{1}\right.  \eqno(99)
\]
\[
\left.  + \frac{1}{m^{2}} D^{1} \frac{1}{\partial^{1}} (D^{1}+m) \frac
{1}{D^{-}} (D^{1}-m) \frac{1}{\partial^{1}} \right] ^{ab} \delta(x^{1} -
y^{1})\delta(x^{-} - y^{-}).
\]
The field $A^{a-}$ occuring explicitly in eq. (99) is dependent on $A^{a1}$
once the constraints $\theta_{a}$ are applied; from eqs. (89a, 90, 94) we find
that
\[
A^{a-} = \left( \frac{1}{D^{1}(D^{1}-m)+m\partial^{1}}\right) ^{ab} \left(
(D^{1}+m)\partial^{-}A^{1}\right) ^{b}.\eqno(100)
\]
We thus see that although the Hamiltonian of eq. (94) is exceedingly simple,
the dynamics is complicated by the Dirac Bracket of eqs. (99, 100).

\section{First and Second Order Forms of the Einstein-Hilbert Action}

The $d$-dimensional Einstein-Hilbert action is
\[
S_{d} = \int d^{d}x \sqrt{-g} \,g^{\mu\nu} \left( \Gamma_{\mu\nu,\lambda
}^{\lambda}- \Gamma_{\lambda\mu,\nu}^{\lambda}+ \Gamma_{\lambda\sigma
}^{\lambda}\Gamma_{\mu\nu}^{\sigma}- \Gamma_{\sigma\mu}^{\lambda}%
\Gamma_{\lambda\nu}^{\sigma}\right) . \eqno(101)
\]
As in refs. [2,3], we assume that $\Gamma_{\mu\nu}^{\lambda}= \Gamma_{\nu\mu
}^{\lambda}$; that there is no torsion in the theory [24]. (Torsion does arise
in supergravity theories because of the coupling to spinor fields.) The affine
connection $\Gamma_{\mu\nu}^{\lambda}$ can either be taken to be an
independent field when $d > 2$ or be identified with the Christoffel symbol
\[
\Gamma_{\mu\nu}^{\lambda}= \left\lbrace
\begin{array}
[c]{c}%
\lambda\\
\mu\nu
\end{array}
\right\rbrace = \frac{1}{2} g^{\lambda\rho} \left(  g_{\mu\rho,\nu} +
g_{\nu\rho,\mu} - g_{\mu\nu,\rho}\right) \eqno(102)
\]
as the equation of motion for $\Gamma_{\mu\nu}^{\lambda}$ has a solution given
by eq. (102) [2, 3].

If $d=2$, then the equation of motion for $\Gamma_{\mu\nu}^{\lambda}$
following from eq. (101) does not have a unique solution; in this case [14,
15]
\[
\Gamma_{\mu\nu}^{\lambda}=\left\{
\begin{array}
[c]{c}%
\lambda\\
\mu\nu
\end{array}
\right\}  +\left(  \delta_{\mu}^{\lambda}K_{\nu}+\delta_{\nu}^{\lambda}K_{\mu
}-g_{\mu\nu}K^{\lambda}\right)  \eqno(103)
\]
where $K^{\lambda}$ is an arbitrary vector. If eq. (103) is used to eliminate
$\Gamma_{\mu\nu}^{\lambda}$ in eq. (101), all dependence on $K^{\lambda}$
cancels. Consequently, in dimensions $d>2$, the first and second order forms
of the EH action are equivalent while if $d=2$ the two forms are non-equivalent.

In order to remove this apparent inequivalence between the first and second
order forms of $S_{2}$, one could enforce eq. (102) by supplementing $S_{2}$
with
\[
S_{w}=\int d^{2}x\,W_{\lambda}^{\mu\nu}\left(  \Gamma_{\mu\nu}^{\lambda
}-\left\{
\begin{array}
[c]{c}%
\lambda\\
\mu\nu
\end{array}
\right\}  \right)  \eqno(104)
\]
where $W_{\lambda}^{\mu\nu}$ is a Lagrange multiplier field. This ensures that
$\Gamma_{\mu\nu}^{\lambda}=\left\{
\begin{array}
[c]{c}%
\lambda\\
\mu\nu
\end{array}
\right\}  $ even when $d=2$. It would be of interest to determine the gauge
symmetries associated with the action of eqs. (101) and (104) that are implied
by the first class constraints that arise.

If we define alternate variables to replace $g_{\mu\nu}$ and $\Gamma_{\mu\nu
}^{\lambda}$,
\[
h^{\mu\nu}=\sqrt{-g}\,g^{\mu\nu},\quad G_{\mu\nu}^{\lambda}=\Gamma_{\mu\nu
}^{\lambda}-\frac{1}{2}\left(  \delta_{\mu}^{\lambda}\Gamma_{\nu\sigma
}^{\sigma}+\delta_{\nu}^{\lambda}\Gamma_{\mu\sigma}^{\sigma}\right)
,\eqno(105a,b)
\]
then eq. (101) becomes
\[
S_{d}=\int d^{d}x\,h^{\mu\nu}\left(  G_{\mu\nu,\lambda}^{\lambda}+\frac
{1}{d-1}G_{\lambda\mu}^{\lambda}G_{\sigma\nu}^{\sigma}-G_{\sigma\mu}^{\lambda
}G_{\lambda\nu}^{\sigma}\right)  .\eqno(106)
\]
The equation of motion for $G_{\mu\nu}^{\lambda}$ that follows from eq. (106)
has the solution when $d>2$
\[
G_{\mu\nu}^{\lambda}=\frac{1}{2}h^{\lambda\rho}\left(  h_{\mu\rho,\nu}%
+h_{\nu\rho,\mu}-h_{\mu\nu,\rho}\right)  -h^{\lambda\rho}h_{\mu\nu}%
\partial_{\rho}\ln\left[  (-h)^{\frac{1}{2(d-2)}}\right]  \eqno(107)
\]
where $h=\det h^{\mu\nu}$. If $d=2$, then the equation of motion for
$G_{\mu\nu}^{\lambda}$ that follows from eq. (106) is consistent only if
$h_{,\lambda}=0$. This is not unexpected, as eq. (105a) shows that
\[
h=-(-g)^{-1+d/2}\eqno(108)
\]
which is constant when $d=2$. Furthermore, when $d=2$ and $h_{,\lambda}=0$,
the equation of motion for $G_{\mu\nu}^{\lambda}$ that follows from eq. (77)
has the unambiguous solution
\[
G_{\mu\nu}^{\lambda}=\frac{1}{2}h^{\lambda\rho}\left(  h_{\mu\rho,\nu}%
+h_{\nu\rho,\mu}-h_{\mu\nu,\rho}\right)  +h_{\mu\nu}X^{\lambda}\eqno(109)
\]
where $X^{\lambda}$ is undefined; it is the analogue of $K^{\lambda}$
occurring in eq. (103). If eq. (109) is used to eliminate $G_{\mu\nu}%
^{\lambda}$ in eq. (77), the $X^{\lambda}$ becomes a Lagrange multiplier that
ensures that $h_{,\lambda}=0$ when $d=2$.

If $d = 2$, the canonical structure of eq. (106) leads to a gauge
transformation that is distinct from diffeomorphism invariance [6-9],
\[
\delta h^{\mu\nu} = \left( \epsilon^{\mu\rho}h^{\nu\sigma} + \epsilon^{\nu
\rho} h^{\mu\sigma}\right) w{_{\rho\sigma}}\;, \quad\delta G_{\mu\nu}%
^{\lambda}= \epsilon^{\lambda\rho}w_{\mu\nu,\rho} + \epsilon^{\rho\sigma
}\left( G_{\mu\rho}^{\lambda}w_{\mu\sigma} + G_{\nu\rho}^{\lambda}w_{\mu
\sigma}\right) .\eqno(110)
\]

\section{Discussion}

In this paper, we have presented a detailed analysis of the canonical
structure of the first order form of the Maxwell, Maxwell-Chern-Simons and
Yang-Mills-Chern-Simons actions. We have also introduced a first order
non-Abelian version of this model. However, more importantly, the procedure
outlined serves as a model for how to perform a fully consistent canonical
analysis of the EH action in General Relativity when it is expressed in first
order form. In refs.\ [6, 7, 8, 9], the Lagrangian $\sqrt{-g}g^{\mu\nu}%
R_{\mu\nu}(\Gamma)$, in so-called Palatini form, in two dimensions is analyzed
using the Dirac constraint formalism employed above. Here we have noted
several aspects of the relationship between the first and second order form of
the EH action for $d=2$.

As has been noted in refs.\ [8, 9], the usual Arnowitt-Deser-Misner approach
[16] to the canonical structure of the EH action involves elimination at the
outset of canonical variables through use of all equations of motion that are
independent of time derivatives, irrespective of whether these equations
correspond to first or second class constraints. (This is most explicitly seen
in the presentation of the first order EH action appearing in ref.\ [17].) We
have circumvented this shortcoming in the analysis of ref. [16] through a
careful application of the Dirac constraint formalism (in which first class
constraints are not used to eliminate dynamical degrees of freedom) to the
first order EH action in dimensions higher than two [18]. The elimination of
fields through use of equations of motion which are independent of time
derivatives was proposed in ref. [17, 25, 42]. This approach is deficient
because if these equations of motion correspond to first class constraints,
then as can be seen from our discussions above one loses a generator of gauge
ransformations [4,5]. Indeed, with the first order EH action in $d>2$
dimensions this is particularly serious as then there would be no tertiary
constraints, while tertiary constraints are necessary contributions to the
generator of gauge transformations in order to have second derivatives of the
gauge functions appear in the gauge transformation of the affine connection
[18]. The possibility of tertiary constraints are also ignored in the general
discussion of first order models in ref. [41]. The techniques used in the
constraint analysis of the EH action in ref. [18] are identical to those
illustrated in our discussion of Maxwell, Chern-Simons and Yang-Mills,
Chern-Simons theory, although the technical difficulties are much more formidable.

It would be interesting to apply the Dirac constraint analysis to the first
order form of the EH action when using light-cone coordinates. The second
order form of the EH action in light-cone coordinates has been examined in
ref. [49].

\section{Acknowledgements}

D. G. C. McKeon is grateful to Roger Macleod for a helpful suggestion. R. N.
Ghalati assisted in early stages of this work.

\end{document}